\documentclass[onecolumn,12pt]{article}
\usepackage{times,comment,overpic,graphicx,amssymb,amsmath,xcolor,multicol,overpic,colortbl,tabularx,tabu,multicol,xr,authblk,abstract,float,multirow,array,booktabs,etoolbox,csquotes}
\usepackage[colorinlistoftodos]{todonotes}
\usepackage{url}
\usepackage[margin=1in]{geometry}

\definecolor{myRed}{rgb}{1,0.8,0.8}
\definecolor{rev}{rgb}{0.9019, 0.0274,  0.1647058}
\definecolor{myGreen}{rgb}{0.4274,   0.7529,   0.28235}
\definecolor{myBlue}{rgb}{0.2588,   0.3098,   0.643137}
\definecolor{myCyan}{rgb}{0.46568627,  0.76372549,  0.81960784}
\definecolor{myMagenta}{rgb}{0.70588,   0.29019,   0.61960}
\definecolor{myYellow}{rgb}{1,1,.1}

\title{
{\large Brief for the Canada House of Commons Study on the Implications of Artificial Intelligence Technologies for the Canadian Labor Force}\\
Generative Artificial Intelligence Shatters Models of AI and Labor
}

\author[1,2,3,*]{Morgan R. Frank}

\affil[1]{Department of Informatics and Networked Systems, University of Pittsburgh, Pittsburgh, PA 15216 USA}
\affil[2]{Digital Economy Lab,  Stanford University, Stanford, CA 94305 USA}
\affil[3]{Connection Science, Massachusetts Institute of Technology, Cambridge, MA, USA}
\affil[*]{To whom correspondence should be addressed. E-mail: mrfrank@pitt.edu}
\date{November 2023}

\begin{document}

\maketitle


\section{Introduction}
Exciting advances in generative artificial intelligence (AI) have sparked concern for jobs, education, productivity~\cite{baily2023machines}, and the future of work.
As with past technologies, generative AI may not lead to mass unemployment.
But, unlike past technologies, generative AI is creative, cognitive, and potentially ubiquitous which makes the usual assumptions of automation predictions ill-suited for today.
Existing projections suggest that generative AI will impact workers in occupations that were previously considered immune to automation.
As AI's full set of capabilities and applications emerge, policy makers should promote workers' career adaptability.
This goal requires improved data on job separations and unemployment by locality and job titles in order to identify early-indicators for the workers facing labor disruption.
Further, prudent policy should incentivize education programs to accommodate learning with AI as a tool while preparing students for the demands of the future of work.

\section{Generative AI Shatters Conventional Wisdom}
Rarely is a single technology so pervasive as to make whole occupations obsolete.
Rather, technology performs specific tasks and impacts the demand for workers who perform those tasks in their job (i.e., skill-biased technological change~\cite{hicks1963theory,violante2008skill,acemoglu2011skills}).
Thus efforts to predict technology-driven job disruptions focus on the variation tasks and activities across occupations~\cite{alabdulkareem2018unpacking,frank2019toward}.
However, knowing both the specific capabilities of technology and the activities of workers across industries, firms, and regions has been a difficult challenge.
Historically, researchers have relied on empirical heuristics that broadly characterize the relationship between classes of workers and the technologies they interact with.
Routine physical workers (e.g., in manufacturing and construction) are thought to be replaced by robotics, while non-routine cognitive workers become more innovative and productive with computing and machine learning.

These heuristics inform predictions of workers' exposure to technology.
For example, creative workers---including artists, graphic designers, and even scientists---are typical examples of non-routine cognitive occupations that have been assumed to be safe from automation.
First appearing as a pre-print in 2013, a high-profile study~\cite{frey2017the} from Oxford University asserted:
\begin{displayquote}
    \emph{
    ``Because creativity, by definition, involves not only novelty but value, and because values are highly variable, it follows that many arguments about creativity are rooted in disagreements about value\dots 
    In the absence of engineering solutions to overcome this problem, it seems unlikely that occupations requiring a high degree of creative intelligence will be automated in the next decades.''
    }
\end{displayquote}
Rather than deviating from these norms, automation studies differ mostly in their modeling of technology's capabilities (e.g., based on expert opinion~\cite{brynjolfsson2018what}, public opinion~\cite{felten2018method}, or patents~\cite{webb2019impact,meindl2021exposure}).

Do these heuristics hold with today's technology?
Or does generative AI shatter conventional wisdom?
While we are still waiting to see the full capabilities and applications of generative AI, large language models (LLMs), such as OpenAI's ChatGPT and Google's Bard, and image generators, such as Midjourney and OpenAI's DALL-E, can draft essays, write computer code, and perform graphic design.
These capabilities are clear examples of what would traditionally be called non-routine cognitive work and suggest that a previously immune class of white collar workers may now compete with automation.
For example, a recent study~\cite{eloundou2023gpts} from OpenAI and the University of Pennsylvania finds that highly-educated U.S. workers in high-earning occupations are most exposed to LLMs (i.e., these occupations require workplace activities that could be performed by LLMs).

Generative AI shatters the conventional wisdom of past automation studies.
For instance, exactly one decade after the quote above, AI is performing creative work in the form of writing, image creation, and software development~\cite{epstein2023art}.
AI has evolved and so too must our approach to economics research, education programs, and labor policy.
Rather than fallible heuristics, a response to AI must be informed by detailed empirics representing workers across the economy.

\section{Improve Data on Negative Labor Outcomes}

\begin{figure}[t]
    \centering
    \includegraphics[width=\textwidth]{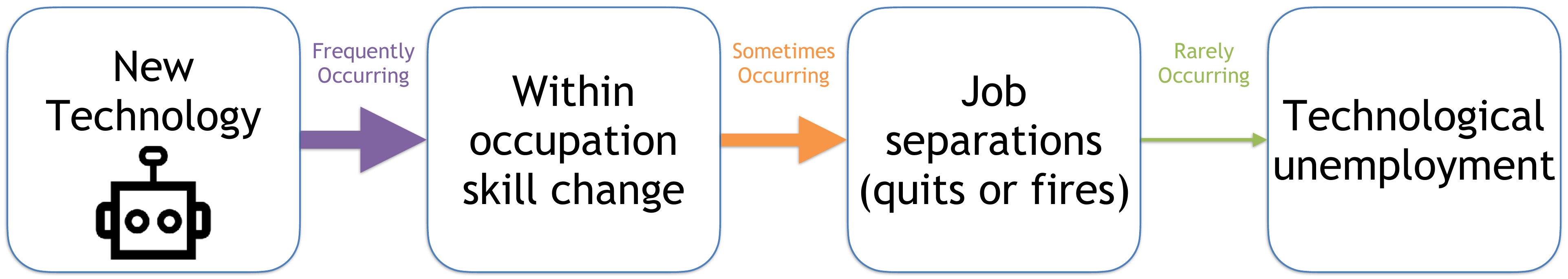}
    \caption{
        The pipeline of technologies' most to least common impact on workers.
        Technology reshapes the demand for specific skills rather than whole occupations.
        If workers fail to adapt their skills, then job separations can occur leading workers to seek new jobs.
        Workers may apply for an receive unemployment benefits if new jobs are not found.
        Data is needed to describe each part of this impact pipeline for each occupation.
    } 
    \label{pipeline}
\end{figure}

New applications of generative AI increase the need for labor data that liberates policy from limiting assumptions regarding automation.
Currently, automation studies compare occupation exposure scores to decreasing employment or wages~\cite{frey2017the,arntz2016risk,brynjolfsson2018what,felten2018method,webb2019impact,tolan2021measuring}, but this approach merely reflects the data that has been available rather than the direct symptoms of AI exposure.
For example, unemployment or job losses associated with an occupation can increase or decrease even as employment for that occupation increases~\cite{mean2023job}. 
Thus, economic policy must be informed by the direct symptoms of workers' exposure to AI.

Generative AI can do certain creative and cognitive activities which may diminish the demand for workers to perform those activities.
Rather than automating an entire occupation, a more direct outcome is that workers shift their activities to complement the work of AI~\cite{frank2019toward} (i.e., within-occupation skill change).
If workers cannot adapt, then job separations (i.e., employer firing or worker quitting) may displace workers~\cite{mortensen1998technological,jovanovic1979job,fossen2022new}.
Displaced workers who cannot quickly find new employment may rely on unemployment benefits while they continue job seeking.
Figure~\ref{pipeline} describes this pipeline of AI impact from the most direct and common symptoms of AI exposure to the least common. 
Note, however, that this pipeline says nothing about employment or wages.

For most national economies, data for describing the pipeline of AI impact does not exist at the level of detail needed to improve automation predictions.
But there are steps to improve the level insights that can inform labor policy in response to generative AI.

{\bf Improved data on within-occupation skill change} must be dynamic enough to model the new skills and capabilities that AI will enable (e.g., prompt engineering: the skill of writing prompts for LLMs) while also subject to regular updates and maintaining representativeness of workplace activities across the economy.
For example, the U.S. Bureau of Labor Statistics (BLS) reports skill and task requirements for over 900 different occupations in the O*NET database;
this database achieves good representation for workers across U.S. sectors but each occupation is only updated every five years and the taxonomy of workplace skills and tasks is updated even less frequently.
With this data alone, it would be difficult for policy makers to respond to emergent demand for new workplace activities, such as ``prompt engineering.''
Researchers have begun to use job postings to supplement federal data, but, currently, job postings can be biased towards certain industries and offer their own data challenges (e.g., processing and cleaning free text job descriptions). 

{\bf Improved data on job separations} requires real-time information on the employers or occupations where job separations are occurring, the nature of the separation (i.e., worker quitting their job or employer firing a worker), and reason for the separation (e.g., worker was not adaptable to new workplace demands versus a human resource issue).
For example, the U.S. BLS Job Openings and Labor Turnover Survey (JOLTS) program produces data on job openings, hires, and separations by month and state, but does not detail the industry, employer, or occupation related to these important labor dynamics.
Canada's provincial economies are diversified enough to require an explanation of job separations in order to better tie job losses to generative AI compared to other sources of disruption.
Online worker resumes (e.g., from LinkedIn or Indeed.com) offer a potential data source with real-time insights into job separations, separated workers' skills, and efforts to seek new employment, but, as with job postings, representativeness and bias towards certain industry may pose a challenge.

{\bf Improved data on unemployment} should go beyond total unemployment by labor market to also include \emph{unemployment risk} by occupation or industry.
Traditionally, ``unemployment by occupation'' is a strange concept because, by definition, unemployed individuals have no occupation.
However, the probability of receiving unemployment by occupation should suffice and, again, enable policy makers and economists to directly tie AI disruptions to unemployment versus other sources.
Typically, unemployment insurance benefits are contingent on recipients' continued effort to seek new employment and, in service to this goal of re-employment, insurance offices often track the employment history of benefit claimants.
Combined with total unemployment statistics and employment shares by occupation, this data could create new insights into the causes of unemployment and portions of the workforce that are in greatest need of policy intervention with the increased adoption of generative AI.

\section{Adapting Education}
Generative AI will reshape the skill requirements of white collar occupations and so the educational pipeline must adapt if today's students are to fill these roles in the future.
As professors and teachers struggle to detect LLM-generated essays in classrooms, a more practical approach should incorporate generative AI applications as learning tools, much like calculators in a mathematics.
It is not that students should completely depend on generative AI, but learning the tools of the day should be part of any curriculum that prepares students for the future of work.

While workforce statistics abound, comparatively fewer economy-wide data reflect the skills taught during higher education and their influence on long-term career outcomes.
For example, the U.S. Department of Education reports enrollment demographics and two-years post-graduation earnings by year, university/college, and major, but offers little insight into career outcomes thereafter or the differences in skills taught within the same major across different universities.
Just as researchers compare technology patents occupations' skills~\cite{webb2019impact,meindl2021exposure}, course descriptions from syllabi may offer a scalable option to infer taught skills from text course descriptions.
In turn, the methods for studying technology in the labor market should be applied to taught skills to identify the majors and universities with the greatest exposure to generative AI. 
These educational pathways will require the most attention as we continue to learn about generative AI's capabilities and applications.
Teachers may focus less on skills that are completely automated by AI.
But, more commonly, skills related to the capabilities of AI will require more attention so that students can complement technology once they enter the workforce.
For example, recent case studies~\cite{dell2023navigating,brynjolfsson2023generative} suggest that generative AI tools do not improve the performance of experts but, instead, raise the performance of non-experts to comparable levels.
If this observation holds across contexts, then incorporating generative AI into learning curricula has the potential to improve learning objectives for under-performing students~\cite{mollick2023assigning,mollick2023using}.

Just as policy makers and government agencies measure skills and labor market outcomes, they should equally measure the sources of labor market skills.
In that case of white collared workers who are most exposed to generative AI~\cite{eloundou2023gpts}, many of these skills result from a college education.
Quantifying the pipeline for specific skills will inform policy to both re-skill today's workers and to prepare today's students for the future of work with generative AI.

\bibliographystyle{unsrt}
\bibliography{main}
\end{document}